\documentclass[a4paper]{report}
\usepackage[utf8]{inputenc}
\usepackage[T1]{fontenc}
\usepackage{RJournal}
\usepackage{amsmath,amssymb,array}
\usepackage{booktabs}

\usepackage{enumitem}

\begin{document}

\sectionhead{Contributed research article}
\volume{XX}
\volnumber{YY}
\year{20ZZ}
\month{AAAA}

\begin{article}
\title{\pkg{npcure}: An R Package for Nonparametric Inference in Mixture Cure Models}
\author{by Ana L\'opez-Cheda, M. Amalia J\'acome, Ignacio L\'opez-de-Ullibarri}

\maketitle

\abstract{
Mixture cure models have been widely used to analyze survival data with a cure fraction. They assume that a subgroup of the individuals under study will never experience the event (cured subjects). So, the goal is twofold: to study both the cure probability and the failure time of the uncured individuals through a proper survival function (latency). The R package \pkg{npcure} implements a completely nonparametric approach for estimating these functions in mixture cure models, considering right-censored survival times. Nonparametric estimators for the cure probability and the latency as functions of a covariate are provided. Bootstrap bandwidth selectors for the estimators are included. The package also implements a non\-pa\-ra\-me\-tric covariate significance test for the cure probability, which can be applied with a continuous, discrete, or qualitative covariate.
}

\section{Introduction}
\label{sec:introduction}
In classical survival analysis, it is assumed that all the individuals will eventually experience the event of interest. However, there are many contexts in which this assumption might not be true. Noticeable examples are the lifetime of cancer patients after treatment, time to infection in a risk population, or time to default in credit scoring, among many others. Cure models are a stream of methods recently developed in survival analysis that take into account the possibility that subjects could never experience the event of interest. See \citet{Maller2} for early references and \citet{Amico} for an updated review.

Let $\mathbf{X}$ be a set of covariates and $Y$ the time to the event of interest with conditional survival function $S\left (t|\mathbf{x}\right )=P\left (Y > t|\mathbf{X}=\mathbf{x} \right)$. Mixture cure models, initially proposed by \citet{Boag}, consider the population as a mixture of two types of subjects: the susceptible of experiencing the event if followed for long enough $\left (Y<\infty \right)$ and the cured ones $\left (Y=\infty \right)$. Hence, the survival function of $Y$ can be written as
\[
S\left (t|\mathbf{x} \right)=1 - p\left (\mathbf{x} \right) + p\left (\mathbf{x} \right) S_0\left (t|\mathbf{x} \right),
\]
where $1-p\left (\mathbf{x} \right)=P\left (Y=\infty | \mathbf{X} = \mathbf{x} \right) = \lim_{t\rightarrow \infty} S\left (t|\mathbf{x} \right)$ is the cure probability, and the (proper) survival function of the uncured subjects or \emph{latency} is $S_0 \left (t|\mathbf{x} \right) = P\left (Y > t|Y<\infty, \mathbf{X}=\mathbf{x} \right)$. A major advantage of these models over the non-mixture approach is that they allow the covariates to have different effect on cured and uncured individuals.

The cure probability, $1-p\left (\mathbf{x} \right)$, is usually estimated parametrically by assuming a logistic form $\log \left (p \left (\mathbf{x} \right )/\left (1-p \left (\mathbf{x}\right ) \right) \right )=\boldsymbol{\beta}^\prime\mathbf{x}$, with $\boldsymbol{\beta}$ a parameter vector. Estimation of $S_0 \left (t|\mathbf{x} \right )$ can be done by assuming a particular parametric distribution for the failure time of the uncured subjects, or more generally, by applying, e.g., proportional hazards (PH) or accelerated failure time (AFT) assumptions. These two approaches lead to parametric \citep[see, e.g.,][]{Farewell1, Farewell2, Denham} or semiparametric \citep[see, e.g.,][]{Kuk, Peng2, Peng1, Li1} mixture cure models.

An attractive feature of parametric and semiparametric models is that they provide close expressions for relevant parameters and functions. On the other hand, the sound inference is guaranteed only if the chosen model fits the data suitably. A problem with these methods is that the parametric assumptions may be incorrect. For example, regarding the cure rate $1-p \left (\mathbf{x} \right)$, there is no reason to believe that the cure rate is monotone in $\mathbf{x}$, let alone that it follows a logistic model. To solve this hassle, \citet{Muller} propose a test statistic to assess whether the cure rate, as a function of $\mathbf{X}$, satisfies a certain parametric model. As for the latency function, $S_0 \left (t|\mathbf{x} \right)$, it is difficult to verify the distributional assumptions of the model. The goodness of fit for the latency function has only been addressed in settings without covariates and in an informal way \citep{Maller2}. The challenge of developing procedures for testing the parametric form of the conditional survival function of the uncured with covariates is even more ambitious. It would lead to curse-of-dimensionality problems and remains an open question.

As a result of the increasing demand for the use of cure models, the number of packages in R accounting for the possibility of cure in survival analysis has grown significantly over the last decade: see the CRAN task view on survival analysis (\url{https://CRAN.R-project.org/view=Survival}). The \CRANpkg{smcure} package \citep{Cai1} fits the semiparametric PH and AFT mixture cure models \citep[see][]{Kalbfleisch}. Besides, the \CRANpkg{NPHMC} package \citep{Cai2} allows to calculate the sample size of a survival trial with or without cure fractions. More recently, the \CRANpkg{flexsurvcure} package \citep{Amdahl} provides flexible parametric mixture and non-mixture cure models for time-to-event data, and the \CRANpkg{rcure} package \citep{Han} incorporates methods related to robust cure models for survival data which include a weakly informative prior in the logistic part. The \CRANpkg{geecure} package \citep{Niu} features the marginal parametric and semiparametric PH mixture cure models for analyzing clustered survival data with a possible cure fraction. Furthermore, the \CRANpkg{miCoPTCM} package \citep{Bertrand} fits semiparametric promotion time cure models with possibly mis-measured covariates, while the \CRANpkg{mixcure} package \citep{Peng3} implements parametric and semiparametric mixture cure models based on existing R packages. For interval-censored data with a cure fraction, the \CRANpkg{GORcure} package \citep{Zhou} implements the generalized odds rate mixture cure model, including the PH mixture cure model and the proportional odds mixture cure model as special cases. The \CRANpkg{intercure} package \citep{Brettas} provides an implementation of semiparametric cure rate estimators for interval-censored data using bounded cumulative hazard and frailty models. 

In contrast with (semi)parametric procedures, nonparametric methods do not rely on data belonging to any particular parametric family or fulfilling any parametric assumption. They estimate the goal functions without making any assumptions about its shape, so they have much wider applicability than alternative parametric methods. A completely nonparametric mixture cure model must consider purely nonparametric estimators for both the cure rate, $1-p(\mathbf{x})$, and latency function, $S_0 \left (t|\mathbf{x} \right)$. Unlike the (semi)parametric approach, nonparametric mixture cure models have been under study only in recent years. \citet{Laska}, building on the Kaplan-Meier (KM) product-limit (PL) estimator of the survival function $S\left (t \right)=P\left (Y>t \right)$ \citep{Kaplan}, derive nonparametric estimators of the cure rate and latency function, but their model does not allow for covariates. More recently, \citet{Xu} propose a nonparametric estimator of the cure rate with one or more covariates, showing its consistency and asymptotic normality. This estimator was further studied by \citet{Lopez1}, who, besides proving that it is the maximum likelihood nonparametric estimator of the cure probability, also obtain an i.i.d.\ representation and proposed a bootstrap-based bandwidth selector. As for the latency function, \citet{Lopez2} introduce a completely nonparametric estimator, studied some theoretical properties, and proposed a bandwidth selector based on the bootstrap.

Although some of the aforementioned packages have a nonparametric flavor, their approach to mixture cure modeling is not completely nonparametric. Our R package \CRANpkg{npcure} \citep{LopezUllibarri} fills the gap by providing implementations of the nonparametric estimator of the cure rate function proposed by \citet{Xu} \citep[further studied by][]{Lopez1} and of the nonparametric estimator of the latency function proposed by \citet{Lopez2}. 

Furthermore, the generalized PL estimator of the conditional survival function, $S\left(t|x\right)$, proposed by \cite{Beran}, is implemented. Note that the estimators of the cure rate and latency implemented in \pkg{npcure} relate strongly to Beran estimator. In any case, Beran estimator is of outstanding importance by its own, as evidenced by the variety of R packages with functions for computing it, like, e.g., \code{Beran()} in package \CRANpkg{condSURV} \citep{MeiraMachado1}, \code{prodlim()} in package \CRANpkg{prodlim} \citep{Gerds} and \code{Beran()} in package \CRANpkg{survidm} \citep{MeiraMachado2}. The function in our package compares advantageously with the aforementioned functions with respect to the issue of bandwidth selection. This smoothing parameter plays an essential role in the bias-variance tradeoff of every nonparametric smoothing method. In \cite{Dabrowska}, an expression for the bandwidth minimizing the asymptotic mean squared error (MSE) of this estimator was obtained, and a plug-in bandwidth selector was proposed based on suitable estimators of the unknown functions in that expression. However, the performance of this bandwidth selector is unsatisfying for small sample sizes, and a cross-validation (CV) procedure is usually preferred \citep[see][among others]{Iglesias, Gannoun}. Recently, \cite{Geerdens} propose an improved CV bandwidth selector, especially with a high censoring rate. To the best of our knowledge, there are not any R packages allowing to compute Beran estimator with a suitable bandwidth selector: while the \pkg{condSURV} and \pkg{survidm} packages do not consider any bandwidth selectors, the \pkg{prodlim} package uses nearest neighborhoods as the smoothing parameter. The \pkg{npcure} package, available from the Comprehensive R Archive Network (CRAN) at \url{https://CRAN.R-project.org/package=npcure}, fulfills this need with the implementation of the CV bandwidth selector for the Beran estimator in \cite{Geerdens}.

In this paper, we explain how the \pkg{npcure} package can be used in the context of nonparametric mixture cure models with right-censored data. The main objective is to estimate the cure probability and latency functions, as well as to perform covariate significance tests for the cure rate. In the next section, we describe our approach to nonparametric estimation in mixture cure models. The methodology applied in the covariate significance tests is presented in another section. Two sections follow, devoted respectively to explain the package functions and to illustrate their use with an application to a medical dataset.

\section{Nonparametric estimation in mixture cure models} 
\label{sec:np_estimation}

One of the specificities of time-to-event data is related to the presence of individuals that have not experienced the event by the end of the study. The observed survival times of these individuals are said to be \emph{right-censored} and underestimate the true unknown time to the occurrence of the event. This situation is usually modeled by considering a censoring variable $C$, with distribution function $G$, which is conditionally independent of $Y$ given the covariate $\mathbf{X}$. The observed data are then $\left \{\left (\mathbf{X}_i,T_i,\delta_i \right) : i=1,\ldots,n\right \}$, where $T=\min \left(Y,C \right)$ is the observed lifetime and $\delta=\mathbf{1}\left(Y\le C \right)$ is the uncensoring indicator. For a one-dimensional continuous covariate $X$, \cite{Xu} propose the following nonparametric kernel-type estimator of the cure rate:
\begin{equation}
  1-\hat p_h \left(x \right)= \prod_{i=1}^{n} \left(1-\frac{\delta_{[i]} B_{h[i]}(x)}{\sum_{r=i}^n B_{h[r]}(x)} \right) = \hat S_h(T_{\max}^1|x),
  \label{eq:incidence}
\end{equation}
where, for $i=1,\ldots,n$, $\delta_{[i]}$ and $X_{[i]}$ are the concomitant status indicator and covariate corresponding to the $i$th ordered time $T_{(i)}$, and
\begin{equation}  \label{NW_weights}
  B_{h[i]}(x) = \frac{K_h \left (x-X_{[i]} \right)}{\sum_{j=1}^n K_h \left (x-X_{[j]} \right)}
\end{equation}
are the Nadaraya-Watson weights, where $K_h(\cdot)=\frac{1}{h} K\left (\frac{\cdot}{h}\right)$ is a rescaled kernel with bandwidth $h \rightarrow 0$. Although some different kernel functions could be considered, the Epanechnikov kernel, defined as
\begin{equation*}
K(u) = \frac{3}{4} (1-u^2) \mathbf{1}(|u| \leq 1),
\end{equation*}
is the one implemented in the \pkg{npcure} package. Moreover, $\hat{S}_h$ is the estimator of the conditional survival function $S$ in \cite{Beran}, and $T_{\max}^1=\max_{\left \{i:\delta_i=1\right \}}T_i$ is the largest uncensored failure time. \cite{Xu} prove the consistency and asymptotic normality of the estimator in (\ref{eq:incidence}), and \cite{Lopez1} show that it is the local maximum likelihood estimator of the cure rate, and obtained an i.i.d. representation and an asymptotic expression for the MSE.

The nonparametric latency estimator proposed by \cite{Lopez1}, and further studied in \cite{Lopez2}, is:
\begin{equation}
   \hat S_{0,b} \left (t|x \right) = \frac{\hat S_{b} \left (t|x \right) - \left(1 - \hat{p}_{b}(x) \right)}{ \hat{p}_{b}\left (x \right)},
   \label{eq:latency}
\end{equation}
where $\hat S_{b}$ is the PL estimator of the conditional survival function $S$ \citep{Beran} and $\hat{p}_{b}$ is the estimator in (\ref{eq:incidence}). As in the case of the cure rate estimator, a smoothing parameter $b$, not necessarily equal to $h$, is needed to compute $\hat S_{0,b}$ in (\ref{eq:latency}).

\subsection{Consistency of the nonparametric estimators} 
\label{sec:existence_of_cure}

The proposed nonparametric estimators of both the cure rate and latency are consistent under the general condition \citep[see][]{Laska, Maller1, Lopez1, Lopez2}
\begin{equation} \label{eq:tau0tauG}
  \tau_0 \leq \tau_G (x),
\end{equation}
where $\tau_0=\sup_x \tau_0(x)$, and $\tau_0(x) = \sup\{ t\geq 0 : S_0\left (t|x \right) > 0 \}$ and $\tau_G(x) = \sup\{ t\geq 0 : G \left(t|x \right) < 1 \}$ are the right endpoints of the support of the conditional distribution of the uncures and the censoring variable, respectively.

The condition in (\ref{eq:tau0tauG}) ensures $1-p\left(x \right)$ and $S_0\left (t|x \right)$ to be consistently estimated when there is zero probability that a susceptible individual survives beyond the largest possible censoring time, $\tau_G \left(x \right)$. Since $T^1_{\max}$ converges to $\tau_0$ in probability \citep[see][]{Xu}, assumption (\ref{eq:tau0tauG}) guarantees that, asymptotically, all times observed after the largest uncensored survival time, $T^1_{\max}$, can be assumed to correspond to cures.

Under condition (\ref{eq:tau0tauG}), $S_0 \left (\tau_{G}(x) |x \right)=0$ and, for large $n$, the cure rate estimator in (\ref{eq:incidence}) tends to a nonparametric estimator of $S \left (\tau_G \left(x \right)|x \right) = 1-p(x) + p(x) S_0\left(\tau_G(x)|x \right) = 1-p(x)$. However, if there could be uncured individuals surviving beyond $\tau_G(x)$, then $S_0\left(\tau_G(x) |x \right) > 0$ and the estimator in (\ref{eq:incidence}) would estimate $S\left(\tau_G(x)|x \right) = 1-p(x) + p(x) S_0\left(\tau_G(x)|x \right) > 1-p(x)$. This might happen, for example, in a clinical trial with fixed maximum follow-up time.

These comments emphasize that care must be exercised in choosing the length of follow-up if cures might be present since too much censoring or insufficient follow-up time could lead to erroneous conclusions. For example, if the last observation is uncensored, then, even if there is considerable late censoring, the estimated cure rate is 0. To avoid these difficulties, particularly with heavy censoring, reasonably long follow-up times and large sample sizes may be required. In this way, $S_0\left(\tau_G (x) |x \right)$ is sufficiently small for the cure rate estimator in (\ref{eq:incidence}) to be close enough to $1-p(x)$.

Thus, when estimating $1-p(x)$ and $S\left (t|x \right)$ for a given $x$ with a data set, it is important to be confident that $\tau_0 \leq \tau_G (x)$. In any case, if the censoring distribution $G(t|x)$ has a heavier tail than $S_0\left (t|x \right)$, the cure rate estimates computed with the nonparametric estimator in (\ref{eq:incidence}) will tend to have smaller biases regardless of the value of $\tau_0(x)$ \citep[see][]{Xu}. \cite{Maller1} propose a simple nonparametric test to assess condition (\ref{eq:tau0tauG}). The procedure is based on the length of the interval $(T^1_{\max},T_{(n)}]$, i.e., the right tail of the KM estimate where it has a constant value. A long plateau with heavy censoring at the right tail of the KM curve is interpreted as evidence that follow-up time has been long enough to conclude that condition (\ref{eq:tau0tauG}) holds.

\subsection{Bandwidth selection}
\label{sec:bandwidth}

The nonparametric estimators in (\ref{eq:incidence}) and (\ref{eq:latency}) depend on two smoothing parameters, $h$ and $b$, respectively. Bootstrap-based selectors for the bandwidth $h$ of the cure rate estimator and the bandwidth $b$ of the latency estimator are proposed by \cite{Lopez1} and \cite{Lopez2}, respectively. The bandwidths are locally chosen so that the selected bandwidths $h_x$ and $b_x$ depend on the point $x$ of estimation. Using locally adaptive bandwidths instead of global ones is advantageous because they adapt to the structure of the underlying function, differentially smoothing its flat and peaky parts.

For a fixed value $x$, the bootstrap bandwidth of the cure estimator, $h_x^*$, was introduced by \cite{Lopez1} as the minimizer of the bootstrap MSE, approximated with $B$ resamples as follows:
\begin{equation} \label{MSE_boot}
MSE_{x}^*(h_x) \simeq \frac{1}{B} \sum_{b=1}^B \left(\hat p_{h_x}^{*b}(x)- \hat p_{g_x}(x) \right)^2,
\end{equation}
where $\hat p_{h_x}^{*b}(x)$ is the estimator of $p(x)$ in (\ref{eq:incidence}) computed with $\left \{\left(X_i^{\ast b},T_i^{\ast b},\delta_i^{\ast b} \right): i=1,\ldots,n\right \}$ (the $b$th bootstrap resample), and using the local bandwidth $h_x$, and $\hat{p}_{g_x}(x)$ is computed with the original sample $\left \{\left(X_i,T_i,\delta_i \right): i=1,\ldots,n\right \}$, and the local pilot bandwidth $g_x$.

With respect to the latency estimator in (\ref{eq:latency}), \cite{Lopez2} propose to choose the bandwidth $b_x$ locally with a bootstrap bandwidth selector. The bootstrap bandwidth of the latency estimator, $b_x^*$, is taken as the minimizer of the bootstrap mean integrated squared error (MISE):
\begin{equation}
MISE_{x}^*(b_x) \simeq \frac{1}{B}\sum_{b=1}^B\int_0^u\left(\hat S_{0,b_x}^{*b}\left(t|x \right) - \hat S_{0,g_x}\left(t|x\right)\right)^2dt,
\label{MISE_boot}
\end{equation}
where $\hat S_{0,b_x}^{*b}\left(t|x \right)$ is the nonparametric estimator of $S_0 \left (t|x \right)$ in (\ref{eq:latency}) computed with the $b$th bootstrap resample and local bandwidth $b_x$, $\hat S_{0,g_x}\left (t|x \right)$ is the same estimator obtained using the original sample and a local pilot bandwidth $g_x$, and $u$ is an adequately chosen upper bound of the integral.

For a fixed covariate value $x$, the procedure for obtaining the bootstrap bandwidth selector of $h_x$ for $\hat{p}_{h_x}(x)$ (respectively, $b_x$ for $\hat S_{0,b_x}\left (t|x \right)$) is as follows:
\begin{enumerate}
\item Generate $B$ bootstrap resamples $\left \{\left (X_i^{\ast b},T_i^{\ast b},\delta_i^{\ast b}\right ): i=1,\ldots,n \right \}$, for $b=1,\ldots,B$.
\item Consider a search grid of bandwidths $h_l \in \left \{h_1,\ldots,h_L\right \}$. For $b=1,\ldots,B$ and $l=1,\ldots,L$, compute the nonparametric estimator $\hat p_{h_l}^{*b}(x)$ (respectively, the nonparametric latency estimator, $\hat S_{0,h_l}^{*b}\left(t|x \right)$) with the $b$th bootstrap resample and bandwidth $h_l$.
\item Compute the nonparametric estimator $\hat p_{g_x}(x)$ (respectively, the nonparametric latency estimator $\hat S_{0,g_x}\left(t|x \right)$) with the original sample and pilot bandwidth $g_x$.
\item For each bandwidth $h_l\in \left \{h_1,\ldots,h_L \right \}$, compute the Monte Carlo approximation of $MSE_{x}^*(h_l)$ in (\ref{MSE_boot}), (respectively, the Monte Carlo approximation of $MISE_{x}^*(h_l)$ in (\ref{MISE_boot})).
\item The bootstrap bandwidth $h_x^*$ for the cure rate estimator (respectively, $b_x^*$ for the latency estimator) is the minimizer of the Monte Carlo approximation of $MSE_{x}^*(h_l)$ (respectively, $MISE_{x}^*(h_l)$) over the grid of bandwidths $\left \{h_1,\ldots,h_L\right \}$.
\end{enumerate}

Following \cite{Lopez1} and \cite{Lopez2}, the bootstrap resamples in Step 1 are generated considering the following procedure, which is equivalent to the simple weighted bootstrap proposed by \cite{Li2} without resampling the covariate $X$:
\begin{enumerate}[label=\Roman*.]
\item Generate $X_1^*,\ldots,X_n^*$ by fixing $X_i^*=X_i$, $i=1,\ldots,n$.
\item For each $i$, compute the weighted empirical distribution $\hat F_{g_{X_i^*}}\left (t,\delta|X_i^* \right)$ with the original sample, where $\hat F_{g_x}\left(t,\delta|x \right)=\sum_{i=1}^n B_{g_xi}(x)\mathbf{1}\left (T_i\leq t, \delta_i \leq \delta \right)$ and $B_{g_xi}(x)$ is computed with a local pilot bandwidth $g_x$ (see (\ref{eq:pilot_g}) below).
\item For each $i$, generate the pair $\left (T_i^*, \delta_i^* \right)$ from the weighted empirical estimator $\hat F_{g_{X_i^*}}\left(t,\delta|X_i^* \right)$ of the conditional distribution.
\end{enumerate}

\cite{Lopez1} and \cite{Lopez2} show that the effect of the pilot bandwidth on the bootstrap bandwidth selectors of $h_x$ and $b_x$ is considerably low. Consequently, the same expression for the pilot bandwidth, $g_x$, is used in Step II of the bootstrap resampling procedure and in the approximation of the $MSE^*_x$ in (\ref{MSE_boot}) for the selection of the bandwidth $h_x$ of the cure rate estimator (respectively, in the approximation of the $MISE^*_x$ in (\ref{MISE_boot}) for the bandwidth $b_x$ of the latency estimator):
\begin{equation}
g_x = \frac{d_k^+(x) + d_k^-(x)}{2} 100^{1/9} n^{-1/9},
\label{eq:pilot_g}
\end{equation}
where $d_k^+(x)$ (respectively, $d_k^-(x)$) is the distance from $x$ to the $k$th nearest neighbor on the right (respectively, on the left). If there are not at least $k$ neighbors on the right (or left), we use $d_k^+(x)=d_k^-(x)$. \cite{Lopez1} show that a good choice for the parameter $k$ is to consider $k = n/4$. The order $n^{-1/9}$ satisfies the conditions in Theorem 1 of \cite{Li2} and coincides with the optimal order for the pilot bandwidth obtained by \cite{Cao} in the case without censoring.

When selecting locally adaptive bandwidths, the results might look a little bit spiky due to its local nature \citep[see, e.g.,][on local bandwidth selection for kernel regression estimators]{Brockmann}. That could be the case for the bootstrap bandwidths for both the cure rate and latency functions. To get rid of the fluctuation of these local bandwidths, $h_x$ and $b_x$ can be further smoothed, for example, by computing a centered moving average of the unsmoothed vector of bandwidths as in \cite{Lopez1}.

\section{Covariate significance tests}
\label{sec:covariate_sign_tests}

In medical studies, it is usually important to assess whether the cure probability depends on a specific covariate, $X$. Noting that the cure rate can be interpreted as the regression function $E\left(\nu|X=x \right)=1-p(x)$, where $\nu$ is the indicator of cure, the question can be cast in the form of a hypothesis test:
\begin{equation}
\begin{cases}
 H_0: E\left(\nu|X\right) = 1-p \\
 H_1: E\left(\nu|X\right) = 1-p(X) \\ 
\end{cases}.
\label{ec:sign_tests_H0H1_Case1}
\end{equation}
Although there are some parametric approaches to deal with this hypothesis testing problem \citep[see][among others]{Muller}, the only completely nonparametric method was introduced by \cite{Lopez3}. Their procedure is based on the test for selecting explanatory variables in nonparametric regression described by \cite{Delgado}. The greatest advantage of the proposed significance test for the cure rate is that although the test is completely nonparametric, no smoothing parameters are required to test (\ref{ec:sign_tests_H0H1_Case1}).

The main challenge when testing (\ref{ec:sign_tests_H0H1_Case1}) is that the cure indicator, $\nu$, is only partially known due to censoring: complete observations are known to be uncured $\left (\nu=0 \right)$, but censored observations might be either cured or uncured (i.e., $\nu$ is unknown). Under right censoring, all of the cured individuals and some of the uncured ones will be censored. This makes it difficult to guess whether a censored observation belongs to the cured or uncured subpopulation. \cite{Lopez3} solved this situation by replacing the unknown and inestimable response variable $\nu$ in (\ref{ec:sign_tests_H0H1_Case1}) by an unknown but estimable response $\eta$ with the same conditional expectation as $\nu$:
\begin{equation}\label{ec:eta}
 \eta=\frac{\nu \left(1-\mathbf{1}(\delta=0,T\leq \tau )\right)}{1- G\left(\tau |X\right)},
\end{equation}
where $\tau$ is an unknown time beyond which a lifetime might be assumed to be cured. \cite{Lopez3} propose to estimate $\eta$ by replacing $G$ and $\tau$ with suitable nonparametric estimators. The censoring distribution is estimated with the generalized PL estimator by \cite{Beran} computed with the cross-validation (CV) bandwidth selector in \cite{Geerdens} when $X$ is continuous and with the stratified KM estimator with the same bandwidth selector otherwise. The cure threshold, $\tau$, is estimated as $\hat\tau=T^1_{\max}$, the largest uncensored observed time. The expression of $\eta$ in (\ref{ec:eta}) avoids the need for an estimator of the unknown cure indicator, $\nu$, since if $\delta_i=1$ or $\left(\delta_i=0, T_i < \hat\tau \right)$ then $\hat\eta_i=0$, whereas if $\left(\delta_i=0, T_i \geq \hat\tau \right)$ then $\hat\eta_i=1/\left(1-\hat G(\hat\tau | X_i) \right)$. It is easy to check that $E\left(\nu|X \right) = E\left(\eta|X \right)$ if the conditional censoring distribution $G\left(t|x \right)$ is independent of the cure status.

Finally, building on \cite{Delgado} and using the estimated values of $\eta$ in (\ref{ec:eta}), the significance test proposed by \cite{Lopez3} is based on the process:
\begin{equation} \label{eq:sign_test}
U_{n}(x)=\frac{1}{n}\sum_{i=1}^{n} \left (\hat{\eta}_{i}- \frac{1}{n}\sum_{j=1}^{n}\hat{\eta}_{j} \right)\mathbf{1}\left( X_{i}\leq x\right ).
\end{equation}
Cram\'er-von Mises (CM) or Kolmogorov-Smirnov (KS) test statistics can be used:
\begin{align}
CM_n &= \sum_{i=1}^n U_n^2(X_i), \nonumber \\
KS_n &= \max_{i=1,\ldots,n} n^{1/2}| U_n(X_i)|.
\end{align}
Note that if $X$ is a nominal variable, it is impossible to compute the indicator function in (\ref{eq:sign_test}). In this case, \cite{Lopez3} propose to consider all the possible `ordered' permutations of the values of $X$ and to compute $U_n(x)$ according to the `ordering' of each permutation. The values of the CM and KS test statistics are given by the maximum of the values $CM_n$ and $KS_n$ computed along with all the permutations.

The distribution of the CM and KS statistics under the null hypothesis is approximated by bootstrap, according to the following steps:
\begin{enumerate}[label=\Alph*.]
\item Obtain $X_i^*$, $i=1,\ldots,n$, by randomly resampling with replacement from $\left \{X_1,\ldots,X_n\right \}$.
\item Estimate the probability of cure under $H_0$ as $1-\hat p=\hat S_n^{KM}\left(T^1_{\max} \right)$, with $\hat S_n^{KM}$ the KM estimator of the survival function $S(t)=P\left(Y>t \right)$. For $i=1,\ldots,n$:
  \begin{enumerate}
  \item[B.1.] Compute $\hat S_{0,b}\left(t|X_i^\ast \right)$, a nonparametric estimator of the latency $S_0\left(t|X_i^\ast \right)$, with the original sample. Set $Y_i^\ast=\infty$ with probability $1-\hat p$, and draw $Y_i^\ast$ from $\hat S_{0,b}(t|X_i^*)$ with probability $\hat p$.
  \item[B.2.] Generate $C_i^\ast$ from a nonparametric estimator of $G\left(t|X_i^\ast \right)$ with the original sample.
  \item[B.3.] Compute $T_i^\ast=\min \left(Y_i^\ast, C_i^\ast \right)$ and $\delta_i^*=\mathbf{1}\left(Y_i^\ast \leq C_i^\ast \right)$.
  \end{enumerate}
\item With the bootstrap resample $\left \{\left(X_i^*, T_i^*, \delta_i^* \right):i=1,\ldots,n \right\}$ compute $\hat\eta_i^*$ for $i=1,\ldots,n$.
\item With $\left \{\left(\hat\eta_i^*, X_i^* \right): i=1,\ldots,n\right \}$, compute the bootstrap versions of $U_n$ in (\ref{eq:sign_test}) and the corresponding CM and KS statistics, $CM_n^*$ and $KS_n^*$.
\item {Repeat Steps A-D above $B$ times in order to generate $B$ values of the CM and KS statistics, $\left \{CM_n^{\ast 1},\ldots, CM_n^{\ast B}\right \}$ and $\left \{KS_n^{\ast 1},\ldots, KS_n^{\ast B}\right \}$.}
\item The $p$-value of the CM (respectively, KS) test is approximated as the proportion of values $\left \{CM_n^{\ast 1},\ldots, CM_n^{\ast B}\right \}$ larger than $CM_n$ (respectively, $\left \{KS_n^{\ast 1},\ldots, KS_n^{\ast B}\right \}$ larger than $KS_n$).
\end{enumerate}

Note that nonparametric estimators of the conditional functions $S_0\left (t|x \right)$ and $G \left(t|x \right)$ are required in Step B. Following \cite{Lopez3}, if $X$ is continuous, then $S_0\left(t|x \right)$ and $G\left(t|x \right)$ are estimated with the nonparametric estimator in (\ref{eq:latency}) and the generalized PL estimator in \cite{Beran}, respectively, and with the corresponding stratified unconditional estimators otherwise.

\section{The \pkg{npcure} package: structure and functionality} 
\label{sec:NPcure}

The \pkg{npcure} package provides several functions to model nonparametrically survival data with a possibility of cure. Table \ref{tab:overview} contains a compact summary of the available functions. The estimators of the cure rate and latency functions, discussed in the section "Nonparametric estimation in mixture cure models", are implemented by \code{probcure()} and \code{latency()}, respectively. The functions \code{probcurehboot()} and \code{latencyhboot()} compute bootstrap bandwidths for these estimators. Another function deserving mention in this context is \code{beran()}, which computes the generalized PL estimator of the conditional survival function $S\left(t|x \right)$. A CV bandwidth for use with \code{beran()} is returned by \code{berancv()}. Given the computational burden of the procedures implemented by the aforementioned functions, all of them make extensive use of compiled C code. The significance test introduced in the previous section is carried out by \code{testcov()}, and \code{testmz()} performs the nonparametric test of \cite{Maller1}. Next, a detailed account of the usage of all these functions is provided.
\begin{table}[h!]
\centering
\begin{tabular}{lp{10.75cm}}
\toprule
Function & Description \\
\midrule
\code{beran} & Computes Beran's estimator of the conditional survival function.\\
\code{berancv} & Computes the CV bandwidth for Beran's estimator of the conditional survival function.\\
\code{controlpars} & Sets the control parameters of the \code{latencyhboot()} and \code{probcurehboot()} functions.\\
\code{hpilot} & Computes pilot bandwidths for the nonparametric estimators of the cure rate and the latency.\\
\code{latency} & Computes the nonparametric estimator of the latency.\\
\code{latencyhboot} & Computes the bootstrap bandwidth for the nonparametric estimator of the latency.\\
\code{print.npcure} & Method of the generic function \code{print} for `npcure' objects.\\
\code{probcure} & Computes the nonparametric estimator of the cure rate.\\
\code{probcurehboot}& Computes the bootstrap bandwidth for the nonparametric estimator of the cure rate.\\
\code{summary.npcure} & Method of the generic function \code{summary} for `npcure' objects.\\
\code{testcov} & Performs covariate significance tests for the cure rate.\\
\code{testmz} & Performs the nonparametric test of \cite{Maller1}.\\
\bottomrule
\end{tabular}
\caption{\label{tab:overview} Summary of the functions in the \pkg{npcure} package.}
\end{table}

The estimation functions in \pkg{npcure} are restricted to one-dimensional continuous covariates. The Epanechnikov kernel is used in the smoothing procedures. Nonparametric estimation with discrete or categorical variables could be dealt with as in other kernel smoothing procedures. A simple approach is to split the sample into a number of subsets according to the covariate values. When the size of the subsamples is not too small, valid unconditional estimates of the cure probability and latency can be computed. Another alternative is the use of special kernels that can handle any covariate types \citep[see][]{Racine}.

Several features are shared by the functions in the package. All functions return an object of S3 class `npcure', formally a list of components. Among these components are the primary outputs of the functions, like the computed estimates for \code{probcure()} and \code{latency()}, the selected bandwidths for \code{probcurehboot()} and \code{latencyhboot()}, or the $p$-values of the tests for \code{testcov()} and \code{testmz()}. The covariate values, observed times, and uncensoring indicators are passed to the functions via the \code{x}, \code{t}, and \code{d} arguments, respectively. Typically, a set of names is passed, which are interpreted as column names of a data frame specified by the \code{dataset} argument. However, \code{dataset} may also be left as \code{NULL}, the default, in which case the objects named in \code{x}, \code{t}, and \code{d} must live in the working directory. More details on these and other arguments are given in the following.

\subsection{Estimation of the cure rate} \label{sec:cure rate estimation}

The estimation of the cure rate using the nonparametric estimator in (\ref{eq:incidence}) is implemented in the \code{probcure()} function:
\begin{example}
probcure(x, t, d, dataset = NULL, x0, h, local = TRUE, conflevel = 0L,
   bootpars = if (conflevel == 0 && !missing(h)) NULL else controlpars())
\end{example}
The \code{x0} argument specifies the covariate values where conditional estimates of the cure rate are to be computed. The bandwidths required by the estimator are passed to the \code{h} argument. The \code{local} argument is a logical value determining whether the bandwidths are interpreted as local (\code{local = TRUE}) or global (\code{local = FALSE}) bandwidths. Notice that if \code{local = TRUE}, then \code{h} and \code{x0} must have the same length. Actually, the \code{h} argument may be missing, in which case the local bootstrap bandwidth computed by the \code{probcurehboot()} function is used. This last function implements the procedure for selecting the bandwidth $h_x^*$ described in the section "Bandwidth selection", and its usage is:

\begin{example}
probcurehboot(x, t, d, dataset, x0, bootpars = controlpars())
\end{example}
The \code{bootpars} argument controls the details of the computation of the bootstrap bandwidth (see section "Bandwidth selection"). In typical use, it is intended to receive the list returned by the \code{controlpars()} function. The components of this list are described in Table~\ref{tab:controlpars}.

\begin{table}[h!]
\centering
\begin{tabular}{lp{11.5cm}}
\toprule
Argument & Description \\
\midrule
\code{B} & Number of bootstrap resamples (by default, \code{999}).\\
\code{hbound} & A vector giving the minimum and maximum, respectively, of the initial grid of bandwidths as multiples of the standardized interquartile range (IQR) of the covariate values (by default, \code{c(0.1, 3)}).\\
\code{hl} & Length of the initial grid of bandwidths (by default, \code{100}).\\
\code{hsave} & A logical specifying if the grid of bandwidths is saved (by default \code{FALSE}).\\
\code{nnfrac} & Fraction of the sample size determining the order $k$ of the nearest neighbor used when computing the pilot bandwidth $g_x$ in (\ref{eq:pilot_g}) (by default, \code{0.25}).\\
\code{fpilot} & Either \code{NULL}, the default, or a function name. If \code{NULL}, the pilot bandwith is computed by the package function \code{hpilot()}. If not \code{NULL}, it is the name of an alternative, user-defined function for computing the pilot.\\
\code{qt} & In bandwidth selection with \code{latencyhboot()}, order of the quantile of the observed times specifying the upper bound of the integral in the computation of the MISE$^*$ in (\ref{MISE_boot}) (by default, \code{0.75}).\\
\code{hsmooth} & Order of a moving average computed to optionally smooth the selected bandwidths. By default is \code{1}, meaning that no smoothing is done.\\
\bottomrule
\end{tabular}
\caption{\label{tab:controlpars} Summary of the arguments of the \code{controlpars()} function.}
\end{table}

The function \code{probcure()} also allows constructing point confidence intervals (CI) for the cure rate. These CIs exploit the asymptotic normality of the estimator \citep{Xu}, using the bootstrap to obtain an estimate of the standard error of the estimated cure rate. The bootstrap resamples are generated by the same procedure described in the section "Bandwidth selection". Denoting by $z_{1-\alpha/2}$ the $1-\alpha/2$ quantile of a standard normal and by $\widehat{se}_B\left(1-\hat{p}_h(x)\right)$ the estimate of the standard error of $1-\hat{p}_h(x)$ with $B$ bootstrap resamples, a $\left(1-\alpha \right)100$\% CI for $1-p(x)$ is computed as:
\begin{equation}
1-\hat{p}_h(x) \mp z_{1-\frac{\alpha}{2}}\widehat{se}_B\left(1-\hat{p}_h(x)\right).
\end{equation}
The confidence level of the CI is specified through the \code{conflevel} argument as a number between 0 and 1. With the special value 0, the default, no CI is computed. Other parameters related to the bootstrap CIs can be passed to the \code{bootpars} argument, typically via the output of the \code{controlpars()} function. These parameters relate to the number of bootstrap resamples and the computation of the pilot bandwidth, and are specified, respectively, by the \code{B} and \code{nnfrac} arguments described in Table \ref{tab:controlpars}.

The usage of these functions is illustrated with a simulated dataset generated from a model where the cure probability is a logistic function of the covariate:

\begin{example}
library("npcure")
n <- 50
x <- runif(n, -2, 2)
y <- rweibull(n, shape = 0.5 * (x + 4), scale = 1)
c <- rexp(n, rate = 1)
p <- exp(2 * x)/(1 + exp(2 * x))
u <- runif(n)
t <- ifelse(u < p, pmin(y, c), c)
d <- ifelse(u < p, ifelse(y < c, 1, 0), 0)
data <- data.frame(x = x, t = t, d = d)
\end{example}

In the next code example, point and 95\% CI estimates of the cure probability are obtained with \code{probcure()} at a grid of covariate values ranging from $-1.5$ to $1.5$. For the estimation, the local bootstrap bandwidths previously computed by \code{probcurehboot()} are passed to the \code{h} argument. The bandwidths, which have been further smoothed with a moving average of 15 bandwidths, are contained in the \code{hsmooth} component of the output of \code{probcurehboot()}. For the bootstrap, 2000 resamples are generated.

\begin{example}
x0 <- seq(-1.5, 1.5, by = 0.1)
hb <- probcurehboot(x, t, d, data, x0 = x0,
   bootpars = controlpars(B = 2000, hsmooth = 15))
q1 <- probcure(x, t, d, data, x0 = x0, h = hb$hsmooth, conflevel = 0.95,
   bootpars = controlpars(B = 2000))
q1
\end{example}
\begin{example}
#> Bandwidth type: local
#>
#> Conditional cure estimate:
#>         h   x0        cure lower 95
#> 0.6212329 -1.5 1.000000000   0.98450759   1.00000000
#> 0.6523881 -1.4 1.000000000   0.87087244   1.00000000
#> 0.6533320 -1.3 1.000000000   0.86080078   1.00000000
#> 0.6606362 -1.2 1.000000000   0.83135572   1.00000000
#> 0.6710717 -1.1 1.000000000   0.82267310   1.00000000
#> 0.6912311 -1.0 0.972213147   0.78259082   1.00000000
#> ...
\end{example}

More compactly, the same bootstrap bandwidths would be selected and the same estimates obtained if \code{h} were left unset when calling \code{probcure()}:

\begin{example}
q2 <- probcure(x, t, d, data, x0 = x0, conflevel = 0.95,
   bootpars = controlpars(B = 2000, hsmooth = 15))
\end{example}

\begin{figure}[htbp]
  \centering
  \includegraphics[width=7.2cm, height=7.2cm]{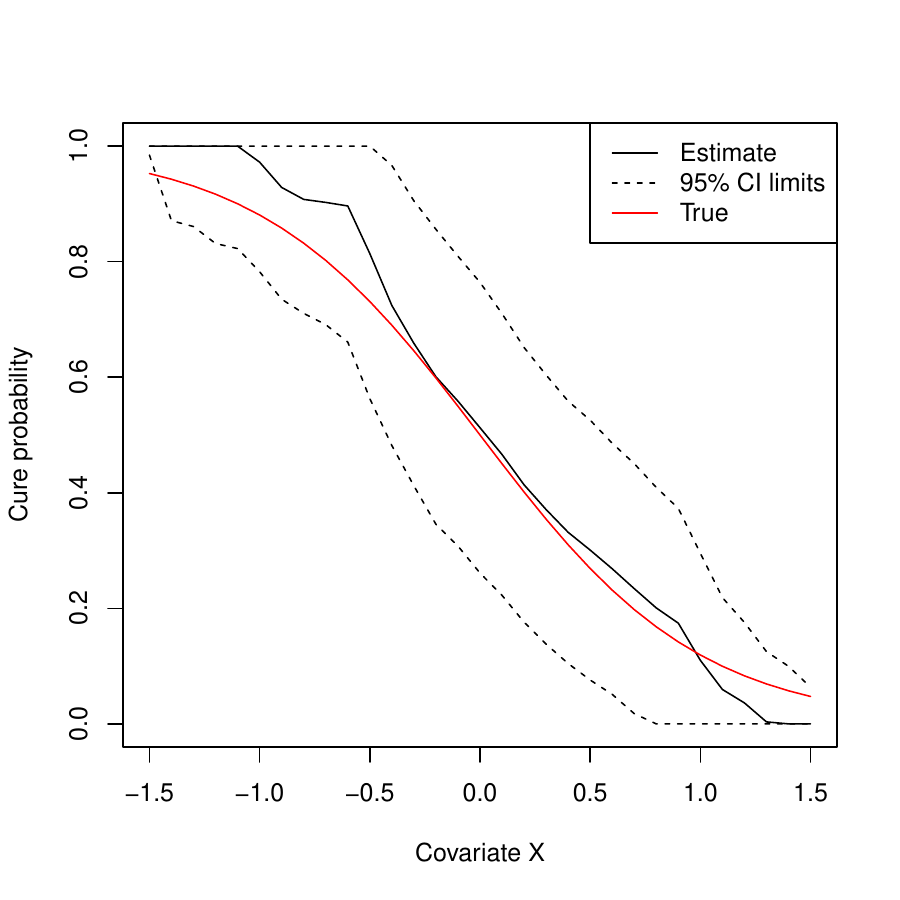}\includegraphics[width=7.2cm, height=7.2cm]{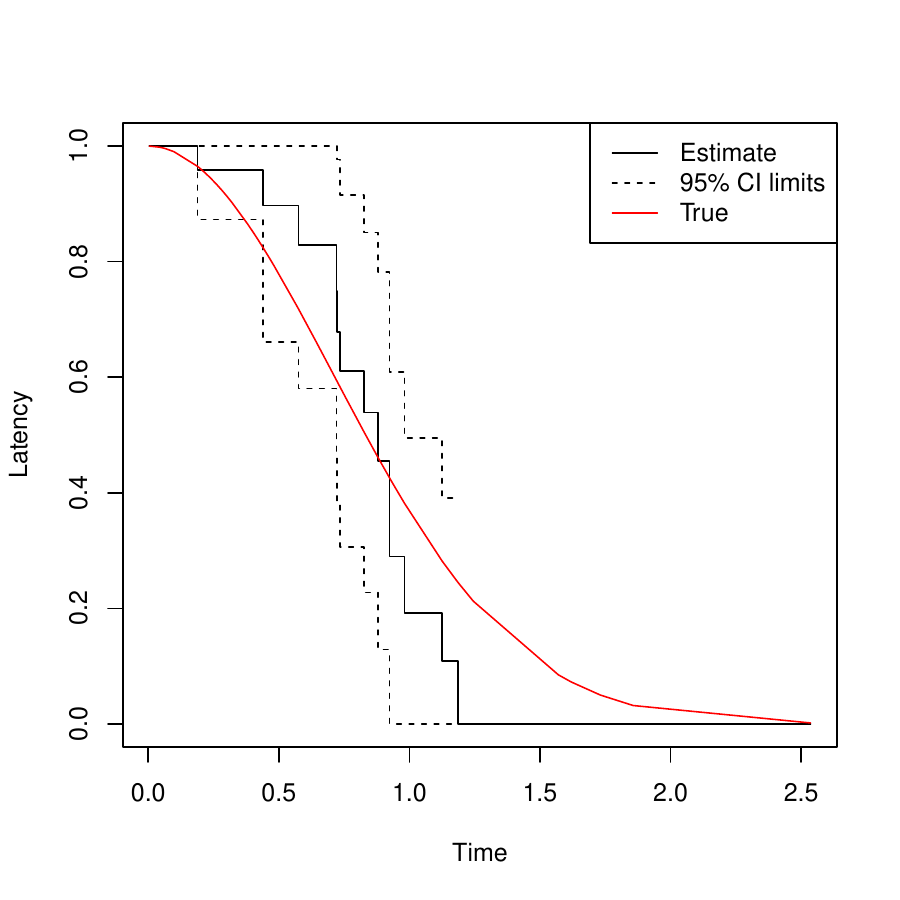}
  \caption{Left panel: estimation of the cure rate. Right panel: estimation of the latency for $x=0$.}
  \label{fig:cure_latency}
\end{figure}

Figure \ref{fig:cure_latency} shows a plot of the true cure rate function and its point and 95\% CI estimates at the covariate values saved in \code{x0}. The plot can be reproduced by executing the next code. The components of the \code{q1} object accessed by the code are \code{x0}, keeping the vector of covariate values, \code{q}, containing the point estimates of the cure rate, and \code{conf}, a list with the lower (component \code{lower}) and upper (component \code{upper}) limits of the CIs for the cure rate.

\begin{example}
plot(q1$x0, q1$q, type = "l", ylim = c(0, 1), xlab = "Covariate X",
   ylab = "Cure probability")
lines(q1$x0, q1$conf$lower, lty = 2)
lines(q1$x0, q1$conf$upper, lty = 2)
lines(q1$x0, 1 - exp(2 * q1$x0)/(1 + exp(2 * q1$x0)), col = 2)
legend("topright", c("Estimate", "95
   lty = c(1, 2, 1), col = c(1, 1, 2))
\end{example}

\subsection{Estimation of the latency function} \label{sec:latency estimation}

The latency estimator in (\ref{eq:latency}) is implemented in the \code{latency()} function:

\begin{example}
latency(x, t, d, dataset = NULL, x0, h, local = TRUE, testimate = NULL,
  conflevel = 0L, bootpars = if (conflevel == 0) NULL else controlpars(),
  save = TRUE)
\end{example}

The function's interface is similar to that of \code{probcure()}, with all the arguments, except for \code{testimate}, having exactly the same interpretation. The \code{testimate} argument determines the times $t$ at which the function $S_0\left(t|x \right)$ is estimated. It defaults to \code{NULL}, which results in the latency being estimated at times given by the \code{t} argument.

Also, as was the case for \code{probcure()}, \code{latency()} allows getting bootstrap CIs for the latency function by specifying their level with the \code{conflevel} argument. These CIs also rely on the asymptotic normality of the latency estimator $\hat S_{0,b} \left(t|x \right)$ in (\ref{eq:latency}) \citep{Lopez2}. A $\left(1-\alpha \right)100$\% CI for $S_{0,b}\left(t|x \right)$ is computed as:
\begin{equation}
\hat S_{0,b} \left(t|x \right) \mp z_{1-\frac{\alpha}{2}}\widehat{se}_B\left(\hat S_{0,b} \left(t|x \right) \right),
\end{equation}
where $\widehat{se}_B\left(\hat S_{0,b} \left(t|x\right)\right)$ is a bootstrap estimate of the standard error of $\hat S_{0,b}\left(t|x \right)$, the bootstrap resamples being generated as described in the section "Bandwidth selection".

Also, as with \code{probcure()}, the user can specify a local or global bandwidth with the combined use of the \code{h} and \code{local} arguments. When \code{h} is left unspecified, a local bootstrap bandwidth is indirectly computed by the \code{latencyhboot()} function:
\begin{example}
latencyhboot(x, t, d, dataset = NULL, x0, bootpars = controlpars())
\end{example}

This function provides an implementation of the bandwidth selector $b_x^*$ introduced in the section "Bandwidth selection". It is homologous to \code{probcurehboot()}, with which it shares a common interface. The only noticeable difference is that now the \code{qt} argument of \code{controlpars()} (see Table \ref{tab:controlpars}) can be used to set $u$, the upper bound of the integral that must be calculated when computing the bootstrap MISE in (\ref{MISE_boot}).

Using the same simulated data as before, the next code illustrates the computation of point and 95\% CI estimates (based on 500 bootstrap resamples) of the latency for covariate values 0 and 0.5, and with local bandwidths equal to 0.8 and 0.5, respectively. Notice that, since the \code{testim} argument is unset, the estimates are computed at the times \code{t}:

\begin{example}
S0 <- latency(x, t, d, data, x0 = c(0, 0.5), h = c(0.8, 0.5),
   conflevel = 0.95, bootpars = controlpars(B = 500))
\end{example}

To estimate the latency using the bootstrap bandwidth selector, \code{latencyhbooot()} can be called before calling \code{latency()}. In the following code, the component \code{h} of the output of \code{latencyhbooot()}, where the selected local bandwidths are contained, is passed to the \code{h} argument of \code{latency()}:

\begin{example}
b <- latencyhboot(x, t, d, data, x0 = c(0, 0.5))
S0 <- latency(x, t, d, data, x0 = c(0, 0.5), h = b$h, conflevel = 0.95)
S0
\end{example}
\begin{example}
#> Bandwidth type: local
#>
#> Covariate (x0): 0.0 0.5
#> Bandwidth (h):  4.531978 2.527206 
#>
#> Conditional latency estimate:
#>
#> x0 = 0 
#>        time   latency lower 95
#> 0.004599127 1.0000000   1.00000000    1.0000000
#> 0.042088293 1.0000000   1.00000000    1.0000000
#> 0.042271452 1.0000000   1.00000000    1.0000000
#> 0.059671372 1.0000000   1.00000000    1.0000000
#> 0.067375891 1.0000000   1.00000000    1.0000000
#> 0.098569312 1.0000000   1.00000000    1.0000000
#> ...
#> 
#> x0 = 0.5 
#>        time   latency lower 95
#> 0.004599127 1.0000000   1.00000000    1.0000000
#> 0.042088293 1.0000000   1.00000000    1.0000000
#> 0.042271452 1.0000000   1.00000000    1.0000000
#> 0.059671372 1.0000000   1.00000000    1.0000000
#> 0.067375891 1.0000000   1.00000000    1.0000000
#> 0.098569312 1.0000000   1.00000000    1.0000000
#> ...
\end{example}

An alternative, more succinct way to proceed is to leave \code{h} unset, since in that case, \code{latencyhboot()} is indirectly called:

\begin{example}
S0 <- latency(x, t, d, data, x0 = c(0, 0.5), conflevel = 0.95)
\end{example}

Figure \ref{fig:cure_latency} shows the estimated and true latencies for covariate value $x=0$. Next, the code to obtain the plot is reproduced, and it is helpful in illustrating the structure of the output list returned by \code{latency()}. The \code{testim} component has the times at which the estimates are computed. The \code{S} component is a list having a named item for each covariate value. Each element contains the latency estimates for a covariate value, and the name is constructed from the covariate value by prefixing it with an \code{x}. The \code{conf} component is also a named list, the names being constructed as those of the \code{S} component. Each one of these items contains, structured as a list, the lower (\code{lower} component) and upper (\code{upper} component) limits of the CIs. Finally, \code{x0} keeps the covariate values as a separate element.

\begin{example}
plot(S0$testim, S0$S$x0, type = "s", xlab = "Time", ylab = "Latency",
   ylim = c(0, 1))
lines(S0$testim, S0$conf$x0$lower, type = "s", lty = 2)
lines(S0$testim, S0$conf$x0$upper, type = "s", lty = 2)
lines(S0$testim, pweibull(S0$testim, shape = 0.5 * (S0$x0[1] + 4), 
   scale = 1, lower.tail = FALSE), col = 2)
legend("topright", c("Estimate", "95
   lty = c(1, 2, 1), col = c(1, 1, 2))
\end{example}

\subsection{Significance test for the cure rate} \label{sec:sig_test}

The \pkg{npcure} package also provides an implementation of the nonparametric covariate significance tests for the cure rate discussed in the section "Covariate significance tests":

\begin{example}
testcov(x, t, d, dataset = NULL, bootpars = controlpars(), save = FALSE)
\end{example}

The \code{x} argument is the covariate whose effect on the cure rate is to be tested. The function's output is a list whose main components are \code{CM} and \code{KS}. Each of them, in turn, is a list containing the test statistic (\code{stat}) and $p$-value (\code{pvalue}) of the CM and KS tests, respectively.

The result of the test carried out with our simulated data and 2500 bootstrap resamples is:

\begin{example}
testcov(x, t, d, data, bootpars = controlpars(B = 2500))
\end{example}
\begin{example}
#> Covariate test 
#>
#> Covariate:  x 
#>               test statistic p.value
#>   Cramer-von Mises 0.4537077  0.0592
#> Kolmogorov-Smirnov 1.2456568  0.0708
\end{example}

Non-numeric covariates can also be tested. For example, for \code{z}, a nominal covariate added to the simulated data, the result is:

\begin{example}
data$z <- rep(factor(letters[1:5]), each = 10)
testcov(z, t, d, data, bootpars = controlpars(B = 2500))
\end{example}
\begin{example}
#> Covariate test 
#>
#> Covariate:  z 
#>               test statistic p.value
#>   Cramer-von Mises 0.2513218  0.6356
#> Kolmogorov-Smirnov 0.7626470  0.5340
\end{example}

\subsection{Estimation of the conditional survival function} \label{sec:survival estimation}

The \pkg{npcure} package also includes the \code{beran()} function, which computes the generalized PL estimator of the conditional survival function, $S\left(t|x\right)$, by \cite{Beran}. The \code{beran()} function in our package may be used together with the \code{berancv()} function:

\begin{example}
berancv(x, t, d, dataset, x0, cvpars = controlpars())
\end{example}

This function computes the local CV bandwidth selector of \cite{Geerdens}. It can be directly called by the user, but in practical work should be more usual an indirect call from the \code{beran()} function, which, as said before, computes the generalized PL estimator of $S\left(t|x \right)$: 

\begin{example}
beran(x, t, d, dataset, x0, h, local = TRUE, testimate = NULL, conflevel = 0L,
   cvbootpars = if (conflevel == 0 && !missing(h)) NULL else controlpars())
\end{example}

The arguments of these two functions have the same meaning as their homonyms in the \code{latency()} and \code{latencyhboot()} functions, \code{cvpars} and \code{cvbootpars} playing the role of \code{bootpars} in these last functions. As in \code{latency()}, if no bandwidth is provided by the user via \code{h}, then the local CV bandwidth in \cite{Geerdens} is computed by \code{berancv()}.

For example, the code below computes the Beran estimator for the covariate values 0 and 0.5 using local CV bandwidths. The default behavior of \code{berancv()} is modified by the auxiliary function \code{controlpars()}. In detail, the local CV bandwidth search is performed in a grid of bandwidths, which is saved (\code{hsave = TRUE}) and consists of 200 bandwidths (\code{hl = 200}) ranging from 0.2 to 2 times the standardized IQR of the covariate (\code{hbound = c(0.2, 2)}). Point and 95\% CI estimates of the conditional survival function $S\left(t|x \right)$ are computed by \code{beran()} with the selected bandwidths:

\begin{example}
x0  <- c(0, 0.5)
hcv <- berancv(x, t, d, data, x0 = x0,
   cvpars = controlpars(hbound = c(0.2, 2), hl = 200, hsave = TRUE))
S <- beran(x, t, d, data, x0 = x0,  h = hcv$h, conflevel = 0.95)
S
\end{example}
\begin{example}
#> Bandwidth type: local
#>
#> Covariate (x0): 0.0 0.5
#> Bandwidth (h):  1.598875 1.104106 
#>
#> Beran's conditional survival estimate:
#>
#> x0 = 0 
#>        time  survival lower 95
#> 0.004599127 1.0000000    1.0000000    1.0000000
#> 0.042088293 1.0000000    1.0000000    1.0000000
#> 0.042271452 1.0000000    1.0000000    1.0000000
#> 0.059671372 1.0000000    1.0000000    1.0000000
#> 0.067375891 1.0000000    1.0000000    1.0000000
#> 0.098569312 1.0000000    1.0000000    1.0000000
#> ...
#>
#> x0 = 0.5 
#>        time  survival lower 95
#> 0.004599127 1.0000000    1.0000000    1.0000000
#> 0.042088293 1.0000000    1.0000000    1.0000000
#> 0.042271452 1.0000000    1.0000000    1.0000000
#> 0.059671372 1.0000000    1.0000000    1.0000000
#> 0.067375891 1.0000000    1.0000000    1.0000000
#> 0.098569312 1.0000000    1.0000000    1.0000000
#> ...
\end{example}

The next code shows an equivalent way of obtaining the same estimates:

\begin{example}
S <- beran(x, t, d, data, x0 = x0, conflevel = 0.95,
   cvbootpars = controlpars(hbound = c(0.2, 2), hl = 200, hsave = TRUE))
\end{example}

\begin{figure}[htbp]
  \centering
  \includegraphics[width=7.2cm, height=7.2cm]{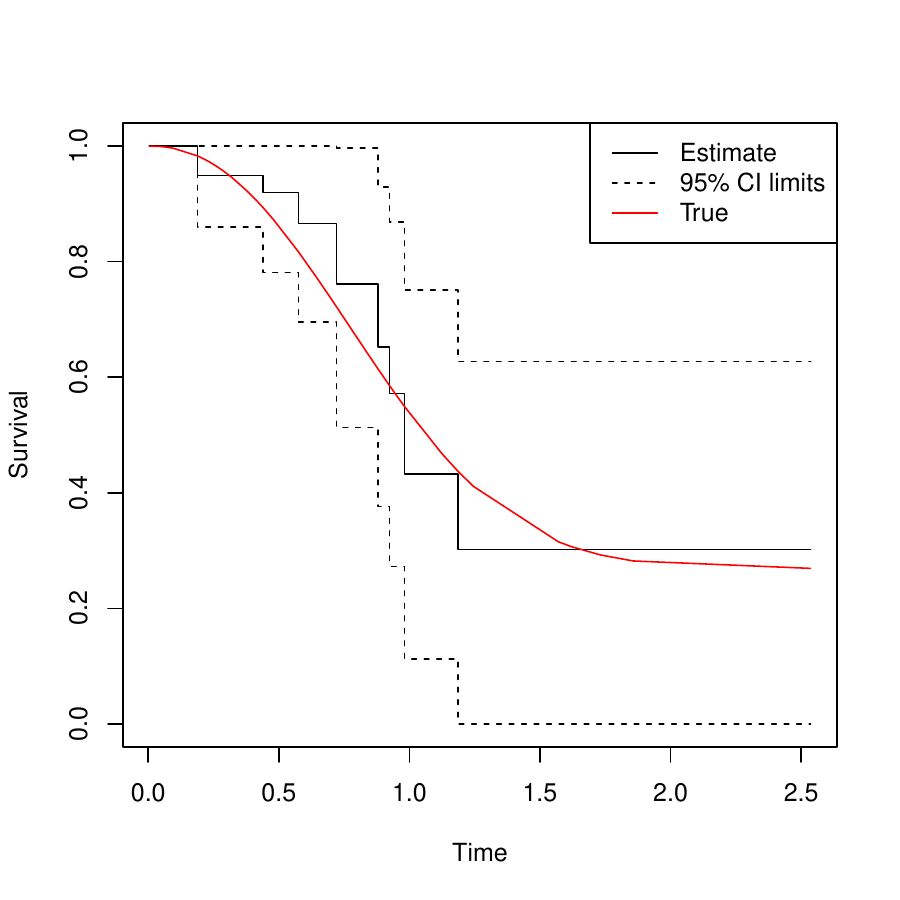}
  \caption{Beran's estimate of the conditional survival function for $x=0.5$.}
  \label{fig:beran} 
\end{figure}

Figure \ref{fig:beran} displays point and 95\% CI estimates of the survival curve for covariate value 0.5. It has been obtained by executing:
\begin{example}
plot(S$testim, S$S$x0.5, type = "s", xlab = "Time", ylab = "Survival",
   ylim = c(0, 1))
lines(S$testim, S$conf$x0.5$lower, type = "s", lty = 2)
lines(S$testim, S$conf$x0.5$upper, type = "s", lty = 2)
p0 <- exp(2 * x0[2])/(1 + exp(2 * x0[2]))
lines(S$testim, 1 - p0 + p0 * pweibull(S$testim,
   shape = 0.5 * (x0[2] + 4), scale = 1, lower.tail = FALSE), col = 2)
legend("topright", c("Estimate", "95
   lty = c(1, 2, 1), col = c(1, 1, 2))
\end{example}

\subsection{Test for enough follow-up} \label{sec:MZ_test}

The nonparametric estimators of the cure rate and latency functions given in (\ref{eq:incidence}) and (\ref{eq:latency}), respectively, require assumption (\ref{eq:tau0tauG}) for their consistency. In other words, the follow-up must be long enough for cures to happen so that the censored times after the largest uncensored observation can be assumed to correspond to cured subjects.

The procedure to test the hypothesis (\ref{eq:tau0tauG}) proposed by \cite{Maller1} is performed by the \code{testmz()} function:

\begin{example}
testmz(t, d, dataset)
\end{example}

The function returns a list (with class attribute `npcure') whose main component, containing the $p$-value of the test, is \code{pvalue}. The further component \code{aux} is, in turn, a list of components \code{statistic}, which contains the test statistic, \code{n}, the sample size, \code{delta}, giving the difference between the largest observed time $T_{(n)}$ and the largest uncensored time $T^1_{\max}$, and \code{interval}, which has the range between $\max (0, T^1_{\max} -$ \code{delta}$)$ and $T^1_{\max}$.

With our simulated data, the result of the test is:

\begin{example}
testmz(t, d, data)
\end{example}
\begin{example}
#> Maller-Zhou test 
#>
#> statistic  n      p.value
#>        43 50 2.024892e-43
\end{example}

\section{Example} 
\label{sec:realdata}

To illustrate the nonparametric modeling of the mixture cure model with the \pkg{npcure} package, we consider the bone marrow transplantation data in \cite{Klein}, available as the \code{bmt} dataset of the R package \CRANpkg{KMsurv} \citep{Klein2}. The data comes from a multi-center study carried out between 1984 and 1989, involving 137 patients with acute myelocytic leukemia (AML) or acute lymphoblastic leukemia (ALL), aged from 7 to 52. Bone marrow transplant (BMT) is the standard treatment for acute leukemia. Transplantation can be considered a failure when leukemia recurs or the patient dies.  Consequently, the failure time is defined as the time (days) to relapse or death. The variables collecting this information are:

\begin{tabular}{ll}
  \code{t2} & Disease-free survival time in days (time to relapse, death, or end of study) \\
  \code{d3} & Disease-free survival indicator (\code{1}: Dead or relapsed, \code{0}: Alive and disease-free) \\
\end{tabular}

The probability of cure after BMT is high, especially if BMT is performed while the patient remains in the chronic phase \citep{Devergie}. Recovery after BMT is a complex process depending on a large set of risk factors, whose status is coded by the following variables:

\begin{tabular}{ll}
  \code{ta} & Time to acute graft-versus-host disease (GVHD).\\
  \code{tc} & Time to chronic GVHD.\\
  \code{tp} & Time to return of platelets to normal levels.\\
  \code{z1} & Patient age (years).\\
  \code{z2} & Donor age (years).\\
  \code{z7} & Waiting time to transplant (days).\\
  \code{group}  & Disease group (\code{1}: ALL, \code{2}: AML low risk, \code{3}: AML high risk).\\
  \code{da} & Acute GVHD indicator (\code{1}: Developed, \code{0}: Never developed).\\
  \code{dc} & Chronic GVHD indicator (\code{1}: Developed, \code{0}: Never developed).\\
  \code{dp} & Platelet recovery indicator (\code{1}: Returned to normal, \code{0}: Never returned to normal).\\
  \code{z3} & Patient gender (\code{1}: Male, \code{0}: Female).\\
  \code{z4} & Donor gender (\code{1}: Male, \code{0}: Female).\\
  \code{z5} & Patient cytomegalovirus (CMV) status (\code{1}: Positive, \code{0}: Negative).\\
  \code{z6} & Donor CMV status (\code{1}: Positive, \code{0}: Negative).\\
  \code{z8} & FAB (\code{1}: FAB grade 4 or 5 and AML, \code{0}: Otherwise).\\
  \code{z9} & Hospital (\code{1}: Ohio State University, \code{2}: Alferd, \code{3}: St. Vincent, \code{4}: Hahnemann).\\
  \code{z10} & Methotrexate (MTX) used for prophylaxis of GVHD (\code{1}: Yes, \code{0}: No).\\
\end{tabular}

\medskip

\begin{figure}[htbp]
  \centering
  \includegraphics[width=7.2cm, height=7.2cm]{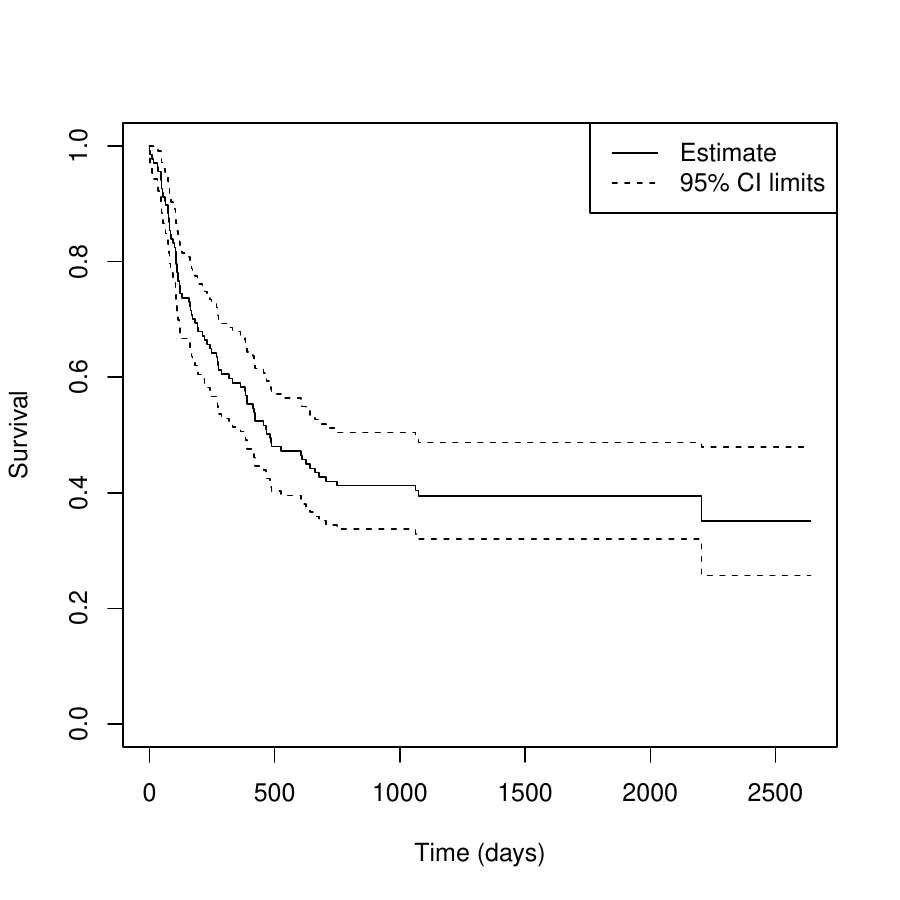}
  \caption{Estimated disease-free survival.}
  \label{fig:KM} 
\end{figure}

Before applying the estimation methods of the \pkg{npcure} package, it should be checked whether the follow-up time was long enough to make it sure that condition (\ref{eq:tau0tauG}) holds. This can be subjectively assessed by visualizing a plot of the KM estimate of the unconditional survival function, $S(t)$. The estimated survival curve in Figure \ref{fig:KM} suggests the existence of a non-zero asymptote at the right tail. The test of \cite{Maller1} confirms that the follow-up period is adequate to ensure the validity of the nonparametric estimation procedures available in the package:
 
\begin{example}
data("bmt", package = "KMsurv")
testmz(t2, d3, bmt)
\end{example}
\begin{example}
#> Maller-Zhou test 
#>
#> statistic   n      p.value
#>        11 137 1.047242e-05
\end{example}

\subsection{Estimation of the probability of cure}

We start by estimating the cure probability as a function of age (\code{z1}) and waiting time to transplant (\code{z7}), respectively. Cure probabilities are estimated at a grid of 100 values between the 5th and 95th quantiles of the values of \code{z1} and \code{z7}. The code for \code{z1} is (for \code{z7}, it is similar):

\begin{example}
x0 <- seq(quantile(bmt$z1, 0.05), quantile(bmt$z1, 0.95), length.out = 100)
q.age <- probcure(z1, t2, d3, bmt, x0 = x0, conflevel = 0.95,
   bootpars = controlpars(hsmooth = 10))
\end{example}
   
\begin{figure}[htbp]
  \centering
  \includegraphics[width=7.2cm, height=7.2cm]{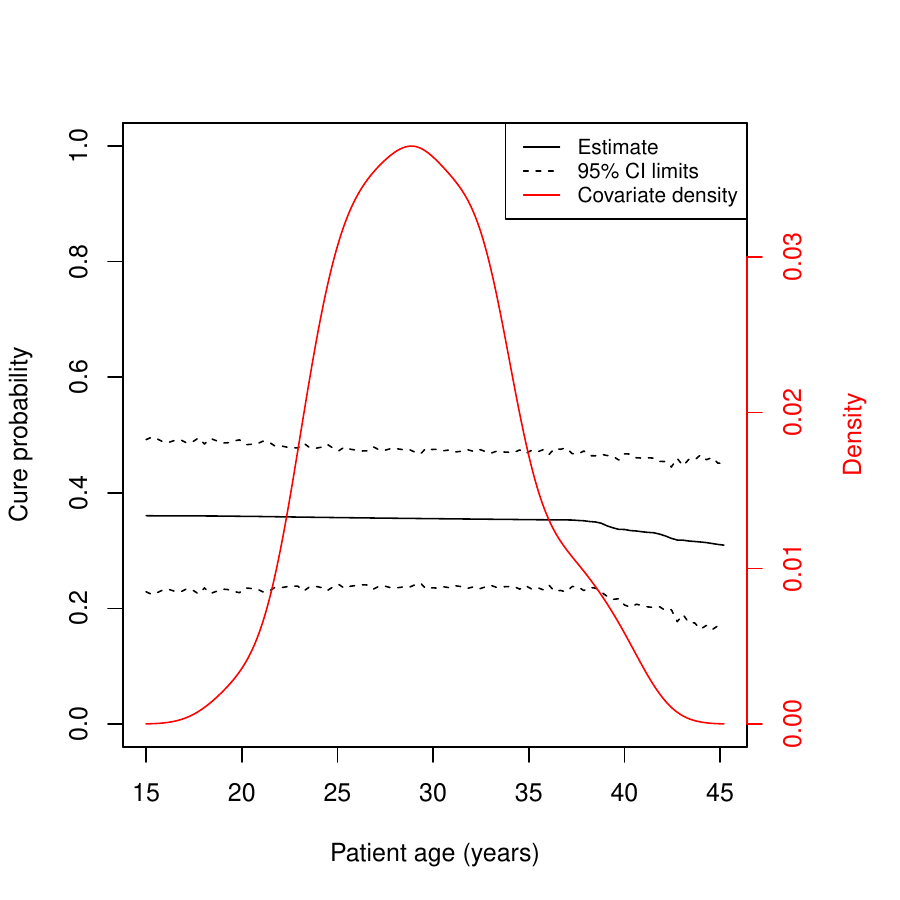}\includegraphics[width=7.2cm, height=7.2cm]{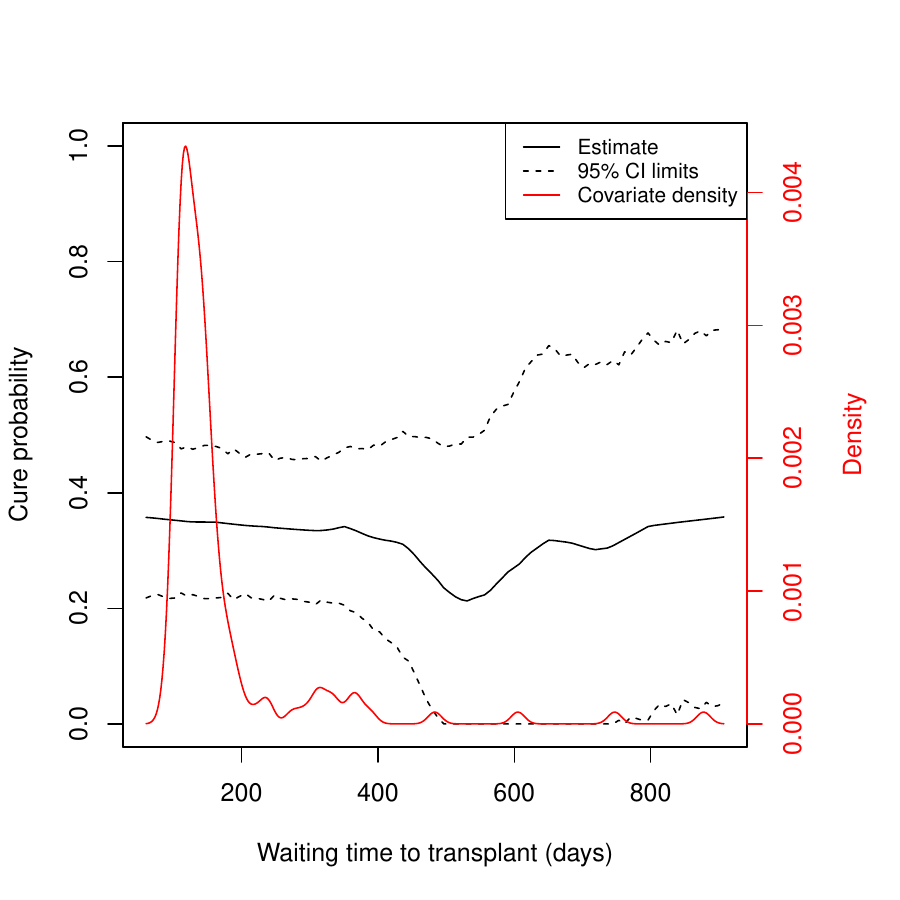}
  \caption{Estimation of the cure probability conditional on age (left panel) and waiting time to transplant (right panel). Nonparametric estimates of the covariate density are plotted for reference.}
  \label{fig:cure} 
\end{figure}

Both estimated cure rates are displayed in Figure \ref{fig:cure}, where a kernel estimate of the covariate density has been added for reference:

\begin{example}
par(mar = c(5, 4, 4, 5) + 0.1)
plot(q.age$x0, q.age$q, type = "l", ylim = c(0, 1),
   xlab = "Patient age (years)", ylab = "Cure probability")
lines(q.age$x0, q.age$conf$lower, lty = 2)
lines(q.age$x0, q.age$conf$upper, lty = 2)
par(new = TRUE)
d.age <- density(bmt$z1)
plot(d.age, xaxt = "n", yaxt = "n", xlab = "", ylab = "", col = 2,
   main = "", zero.line = FALSE)
mtext("Density", side = 4, col = 2, line = 3) 
axis(4, ylim = c(0, max(d.age$y)), col = 2, col.axis = 2)
legend("topright", c("Estimate", "95
   lty = c(1, 2, 1), col = c(1, 1, 2), cex = 0.8)
\end{example}

The cure probability seems to be nearly constant or, at most, to decrease slightly with patient age and as the waiting time to transplant increases.

\subsection{Testing the effect of one covariate on the probability of cure}

The significance of the effect of patient age (\code{z1}) and waiting time to transplant (\code{z7}) on the probability of cure can be tested with the \code{testcov()} function:

\begin{example}
testcov(z1, t2, d3, bmt,  bootpars = controlpars(B = 2500))
\end{example}
\begin{example}
#> Covariate test 
#>
#> Covariate:  z1 
#>               test statistic p.value
#>   Cramer-von Mises 0.1103200  0.8204
#> Kolmogorov-Smirnov 0.7308477  0.7900
\end{example}

\begin{example}
testcov(z7, t2, d3, bmt,  bootpars = controlpars(B = 2500))
\end{example}
\begin{example}
#> Covariate test 
#>
#> Covariate:  z7 
#>               test statistic p.value
#>   Cramer-von Mises 0.7921912  0.0968
#> Kolmogorov-Smirnov 1.6116129  0.1008
\end{example}

The effect of age on the cure probability is not statistically significant with neither the CM nor the KS tests ($p_{CM}$ = 0.820 and $p_{KS}$ = 0.790, where the subscripts identify the $p$-value in an obvious way). As for the effect of waiting time to transplant, it reaches a borderline significance ($p_{CM}$ = 0.097 and $p_{KS}$ = 0.101).

Cure probability can also be compared between groups defined by a categorical covariate. We illustrate this case by considering gender (\code{z3}) and the use of MTX for prophylaxis of GVHD (\code{z10}). For improving readability, we first label the groups:

\begin{example}
bmt$z3 <- factor(bmt$z3, labels = c("Male", "Female"))
bmt$z10 <- factor(bmt$z10, labels = c("MTX", "No MTX"))
summary(bmt[, c("z3", "z10")]) 
\end{example}
\begin{example}
#>      z3         z10    
#> Male  :57   MTX   :97  
#> Female:80   No MTX:40  
\end{example}

\begin{figure}[htbp]
\centering
  \includegraphics[width=7.2cm, height=7.2cm]{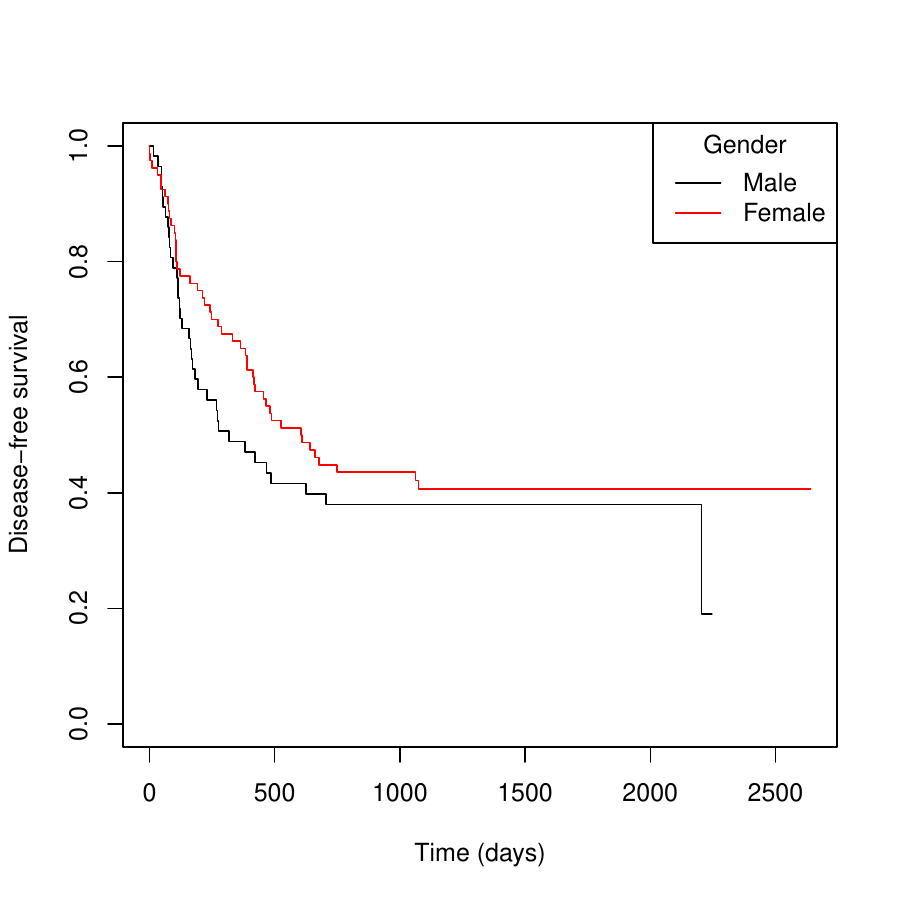}\includegraphics[width=7.2cm, height=7.2cm]{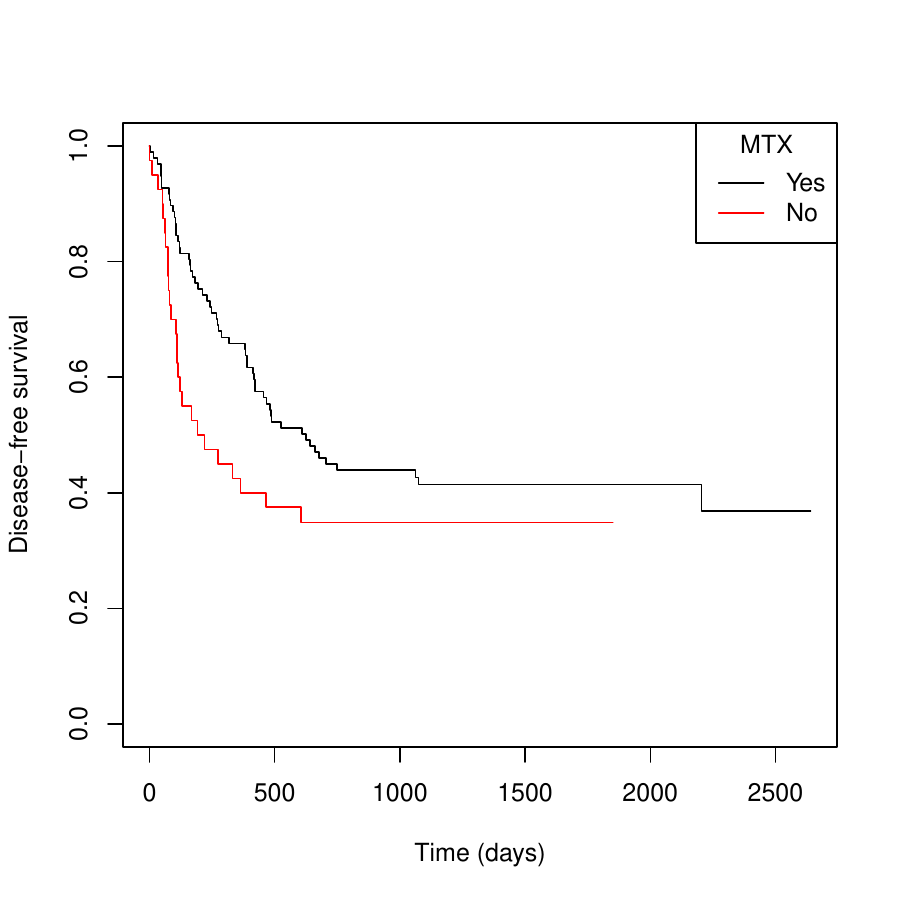}
  \caption{Survival curves of patients conditional on gender (left panel) and use of MTX for prophylaxis of GVHD (right panel).}
  \label{fig:KMcond} 
\end{figure}

The estimated survival functions are displayed in Figure \ref{fig:KMcond}. The code for gender (\code{z3}) is:

\begin{example}
library("survival")
Sgender <- survfit(Surv(t2, d3) ~ z3, data = bmt)
Sgender
\end{example}
\begin{example}
#> Call: survfit(formula = Surv(t2, d3) ~ z3, data = bmt)
#>
#>           n events median 0.95LCL 0.95UCL
#> z3=Male   57     36    318     172      NA
#> z3=Female 80     47    606     418      NA
\end{example}
\begin{example}
plot(Sgender, col = 1:2, mark.time = FALSE, xlab = "Time (days)",
   ylab = "Disease-free survival")
legend("topright", legend = c("Male", "Female"), title = "Gender",
   lty = 1, col = 1:2)
\end{example}

The estimated probability of cure for each group defined by gender (\code{z3}) is obtained by computing for each stratum the unconditional cure rate estimator of \cite{Laska}. This estimator of the probability of cure is the value of the KM curve at $T^1_{\max}$ (i.e., it is the minimum of the KM estimate):

\begin{example}
qgender <- c(min(Sgender[1]$surv), min(Sgender[2]$surv))
qgender
\end{example}
\begin{example}
#> [1] 0.1899671 0.4065833
\end{example}

The estimated probability of cure is 19.0\% for males and 40.7\% for females. The cure probabilities according to the use or not of MTX as GVHD prophylactic (\code{z10}) are:

\begin{example}
Smtx <- survfit(Surv(t2, d3) ~ z10, data = bmt)
qmtx <- c(min(Smtx[1]$surv), min(Smtx[2]$surv))
qmtx
\end{example}
\begin{example}
#> [1] 0.3679977 0.3482143
\end{example}

The cure rate of patients treated with MTX is estimated to be 36.8\%, slightly higher than 34.8\%, the estimate for patients not treated with MTX.

The effect of these two binary variables on the cure probability is tested with the \code{testcov()} function similarly as it was done with continuous covariates:

\begin{example}
testcov(z3, t2, d3, bmt, bootpars = controlpars(B = 2500))
\end{example}
\begin{example}
#> Covariate test 
#>
#> Covariate:  z3 
#>                test statistic p.value
#>   Cramer-von Mises 0.5947305  0.0900
#> Kolmogorov-Smirnov 1.1955919  0.0892
\end{example}
\begin{example}
testcov(z10, t2, d3, bmt, bootpars = controlpars(B = 2500))
\end{example}
\begin{example}
#> Covariate test 
#>
#> Covariate:  z10 
#>               test statistic p.value
#>   Cramer-von Mises  1.018441  0.0692
#> Kolmogorov-Smirnov  1.199340  0.0668
\end{example}

The differences in the probability of cure between males and females, and between patients with and without MTX treatment are not statistically significant, although a borderline effect is evidenced ($p_{CM} = 0.090$ and $p_{KS} = 0.089$ for gender, $p_{CM} = 0.069$ and $p_{KS} = 0.067$ for MTX).

\subsection{Estimation of the latency function}

The survival of the uncured patients (latency) is estimated for patient age (\code{z1}) 25 and 40 years as follows:
\begin{example}
S0 <- latency(z1, t2, d3, bmt, x0 = c(25, 40), conflevel = 0.95,
   bootpars = controlpars(B = 500))
\end{example}
   
\begin{figure}[htbp]
  \centering
  \includegraphics[width=7.2cm, height=7.2cm]{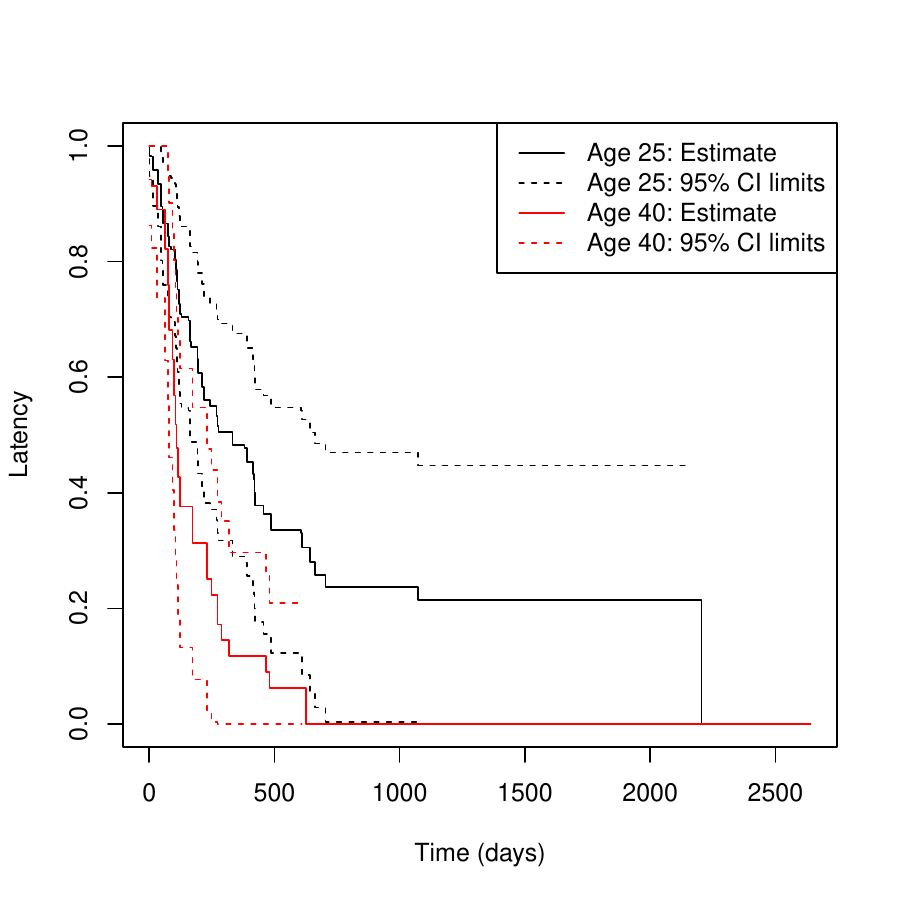}
  \caption{Latency curves of uncured patients 25 and 40 years old.}
  \label{fig:latencycond} 
\end{figure}

Figure \ref{fig:latencycond} displays the survival functions for the two ages, obtained by executing: 
\begin{example}
plot(S0$testim, S0$S$x25, type = "s", ylim = c(0, 1),
   xlab = "Time (days)", ylab = "Latency")
lines(S0$testim, S0$conf$x25$lower, type = "s", lty = 2)
lines(S0$testim, S0$conf$x25$upper, type = "s", lty = 2)
lines(S0$testim, S0$S$x40, type = "s", col = 2)
lines(S0$testim, S0$conf$x40$lower, type = "s", lty = 2, col = 2)
lines(S0$testim, S0$conf$x40$upper, type = "s", lty = 2, col = 2)
legend("topright", c("Age 25: Estimate", "Age 25: 95
   "Age 40: Estimate", "Age 40: 95
   col = c(1, 1, 2, 2))
\end{example}

An increased survival of younger patients can be observed, but the survival advantage vanishes after approximately 6 years.

\section{Summary}

This paper introduces the \pkg{npcure} package. It provides an R implementation of a completely nonparametric approach for estimation in mixture cure models, along with a nonparametric covariate significance test for the cure probability. Moreover, the generalized PL estimator of the conditional survival function with a CV bandwidth selection function is included. Furthermore, the theory underlying the implemented me\-thods, presented in \cite{Xu}, \cite{Lopez1}, and \cite{Lopez2}, has been compiled. 

The \pkg{npcure} package has some limitations. Firstly, it only handles right-censored survival times. Left-censored data, truncation, or interval-censored data have not been considered in this approach, and it remains an open problem to be dealt with in the future. Secondly, a conditional estimation can be performed when only one covariate is involved. The same restriction applies to the implemented covariate significance test for the cure rate. An important extension would be the development of estimation and test procedures for the cure rate and latency functions when they depend on a set of covariates. A major challenge is the way the covariates are handled. In that case, the analysis of a large number of covariates would suffer from the curse of dimensionality. Dimension reduction techniques would be required, which leads to a demanding approach that has not been addressed yet, and we leave for further research.

There is an interesting issue that remains an open problem to be dealt with in future versions of the package. Traditional cure rate models implicitly assume that there is no additional information on the cure status of the patients. So, the cure indicator is modeled as a latent variable. However, examples contradicting this assumption can be found. For instance, in some clinical settings, subjects who are followed up beyond a threshold period without experiencing the event can be considered as cured. In other cases, complementary diagnostic tests providing further information about a patient's cure status may be available. We aim to develop improved non-parametric methods of estimation and hypothesis testing that take into account this additional information.

\section{Acknowledgments}
The first author's research was sponsored by the Beatriz Galindo Junior Spanish Grant from Ministerio de Ciencia, Innovación y Universidades (MICINN) with reference BGP18/00154. All the authors acknowledge partial support by the MICINN Grant MTM2017-82724-R (EU ERDF support included), and by Xunta de Galicia (Centro Singular de Investigaci\'on de Galicia accreditation ED431G/01 2016-2019 and Grupos de Referencia Competitiva CN2012/130 and ED431C2016-015) and the European Union (European Regional Development Fund - ERDF).

\bibliography{lopezcheda}

\address{Ana L\'opez-Cheda\\
  Research Group MODES, CITIC, Departamento de Matem\'aticas, Facultade de Inform\'atica, Universidade da Coru\~na\\
  CITIC, Campus de Elvi\~na s/n, A Coru\~na 15071 \\
  Spain \\
  (ORCiD: 0000-0002-3618-3246)\\
  \email{ana.lopez.cheda@udc.es}}

\address{M. Amalia J\'acome \\
  Research Group MODES, CITIC, Departamento de Matem\'aticas, Facultade de Ciencias, Universidade da Coru\~na\\
  R\'ua da Fraga s/n, A Zapateira, A Coru\~na 15071  \\
  Spain \\
  (ORCiD: 0000-0001-7000-9623)\\
  \email{maria.amalia.jacome@udc.es}}

\address{Ignacio L\'opez-de-Ullibarri \\
  Research Group MODES, Departamento de Matem\'aticas, Escuela Universitaria Polit\'ecnica, Universidade da Coru\~na\\
  15405, Ferrol, A Coru\~na\\
  Spain \\
  (ORCiD: 0000-0002-3438-6621)\\
  \email{ignacio.lopezdeullibarri@udc.es}}

\end{article}

\end{document}